\documentclass[letterpaper,12pt]{article}
\usepackage[russian,english]{babel}
\usepackage {graphicx}
\usepackage{indentfirst}

\begin{document}
\title{THE CONCEPT OF AN ORDER AND ITS APPLICATION FOR RESEARCH
 OF THE DETERMINISTIC CHAINS OF SYMBOLS}
\author{ Ilyevsky V.I.}

\maketitle
\begin{abstract}
The present work is dedicated to searching parameters, alternative
to entropy, applicable for description of highly organized systems.
The general concept has been offered, in which the system complexity
and order are functions of the order establishment rules. The
concept of order poles has been introduced. The concept is being
applied to definition of the order parameter (OP) for non-random
sequences with equal number of zeros and ones. Properties of the OP
are being studied. Definition of the OP is being compared to
classical definition of amount of information.

\end{abstract}
\frenchspacing

\section{The complexity and order concept}
The determined chain term means the readable object that gives the
same sequence of characters for any rereading. These chains are, for
example, DNA, RNA, proteins, any book on the shelf. The issue of how
to determine degree of order and amount of information in this
system is under discussion since publication of Shannon's ~\cite
{art1} works. The Shannon's entropy is the parameter that describes
uncertainty of the system. Information is regarded as elimination of
uncertainty and is calculated provided probabilities of occurrence
of symbols from a given alphabet are defined ~\cite {art2}. Applying
entropy to determined sequences we simulate it by random sequence of
characters. In 1965, A.N. Kolmogorov ~\cite {art3} turned his
attention to limitations of the probabilistic approach in
determining the amount of information and offered an algorithmic
approach, known today as the Kolmogorov's complexity. The
Kolmogorov's complexity does not depend on any probabilistic
assumptions, but the sequence-generating algorithm.

The foregoing does not detract from the importance of the
probabilistic method, in particular, for statistical processing of
experimental data in molecular biology. In this way, in works ~\cite
{art4,art5,art6} the Logo for certain DNA sections is built based on
the Shannon's information.

However, it is very problematic to perceive information contained in
biological systems as eliminated uncertainty. Describing the living
matter as a complex highly organized system, we nevertheless know
~\cite {art7} that entropy of a piece of rock is not much higher
than that of living creatures of the same mass. In quest for other
definitions of complexity and order, we offer the following concept.

Consider a system described by a set of discrete parameters that
determines its state. The system can be a real physical or abstract.
Let us say the system has a complexity if:

\begin{itemize}
\item
The external system, which, affecting this system according to
certain rules or natural laws, transforms it from one state to
another, can be built (defined).

\item
A positive value $T$, which describes external influence, resulting
in the $A\to B$ transition, is defined. Let us regard minimum value
of $T$ as complexity of the $A\to B$ transition. For example,
complexity for the physical systems can be minimal energy or time
necessary for transition from one state to another.

\end{itemize}
We shall establish that a system may be considered ordered provided
it is possible to define specific states, known as poles, defined by
the following:

\begin{itemize}
\item
There is a parameter describing the system, which reaches its
extremes at the poles.
\item Complexity of transition between the poles is maximal.

\end{itemize}

In ordered systems, degree of the state order $A$ can be described
as a function of transition complexities from state $A$ to the order
poles. In the above concept, complexity of transitions and the
system's order states depend on ordering of the external influence.
Such definition may seem very strange.  Let us recall, however, the
assertion "living  comes from the living only" that nobody has
neither proven no refuted as yet. For our purposes, this statement
can be formalized as follows:

 \emph{The system can be ordered only with another ordered system's help.}

Now one may easily agree that degree of the system's order depends
on who performs the procedure and how.

The order poles can be seen in simple physical systems. Let there be
a mixture of solid and liquid phase at the phase transition
temperature. Parameters, describing the system, are masses of $m_1$
and $m_2$ phases, which are, generally speaking, discrete. The
system is closed - the $M=m_1+m_2$ mass is constant. Let us build an
external system - the calorimeter, allowing for heat transfer.
Energy, transferred to the system, changes both $m_1$ and $m_2$, and
its absolute value may be considered the transition complexity.
Maximum complexity is achieved upon complete transition from one
phase to another. Thus, there are two state poles. Another physical
example is an atom in different energy states. The order poles in
this case - the ground state and the ion. In systems, described by
multiple parameters, there can be a large number of poles. The order
poles concept may be of  interest when studying biological systems.
A very general idea, described in the present study, is realized to
determine the order parameter for a finite sequence of symbols with
equal numbers of ones and zeros.

\section{The minimum information poles \\ in the classical theory}

Since the order poles concept is central to this paper, we consider
it necessary to present preliminary researches and show that within
the framework of classic information theory it is possible to define
the poles as states of perfect order with minimum information.

A remark from work ~\cite {art8} brings us to the idea of zero
information determination: "DNA can not be periodic, just as any
sequence of letters in a meaningful  textbook". Intuitive assertion
of minimum information  for a periodic chain can actually be
obtained  by strict  formulation  of the problem.

 Let us examine long chains, composed of zeros and ones, assuming it is possible to
calculate probabilities of character appearance using a statistical
method. Based on these probabilities we can calculate the amount of
information  that a chain of known length contains. We restrict
ourselves to Markov's chains of the first order.

First, let us note that the ideal periodic sequence of symbols can
be simulated in the Shannon's classic  model. Of course, information
in a periodic chain of two characters with a known first character
equals zero. However, it is important to figure out behavior of the
Shannon's information in case of a small deviation from the chain
periodicity. Besides, for comparison with the order parameter we
shall need the maximum information condition for the Markov's chain.

Let us consider a sequence in which $0$ and $1$ occurs with equal
probabilities. Let us further suppose that in the $2n$ symbols long
chain there are on average $k$ pairs of $00$ sequences. (To
calculate pairs without restricting generality, we can close the
chain  into the  ring. If $s\geq2$ zeros follow in sequence, we
believe there are $s-1$ pairs of $00$. It is then easy to calculate
~\cite {art2} the amount of information in such $2n$ symbol long
chain:

\begin{equation}
\label{H2n} H(k)=1-\frac{2n-1}{n}\lbrack
k\log_2\frac{k}{n}+(n-k)\log_2\frac{n-k}{n}\rbrack
\end{equation}

Analyzing function (\ref{H2n}), let us write down the extreme values
of $H(k)$ at  $n\gg1$, as well as values in the field $0\leq k<n$,
immediately following the minimum in the ascending order.

\begin{equation}
\label{Hmax}H_{max}=H(n/2)=2n\, bit
\end{equation}
\begin{equation}
\label{Hmin}H_{min}=H(0)=1\, bit
\end{equation}
\begin{equation}
\label{H1}H_1=H(1)=2\log_2n \,bit
\end{equation}

\begin{equation}
\label{H2}H_2=H(n-1)=2\log_2n\, bit
\end{equation}

We get the maximum amount of information when pairs of $00$ and $11$
appear with the same frequency as pairs of $01$ or $10$. This
situation corresponds to a chain without memory.

 Let us consider the situation near information minimum. At $k=0$,  we have a
deterministic periodic chain (hereinafter $S_+$) with period $01$,
starting either from zero or one:

\begin{equation}
\label{0101} 0101010101...
\end{equation}

The amount of information in a periodic chain of $2n$ symbols equals
the one carried by the first symbol. If the first character is
known, then, as we have already noted, information equals zero.

 At $k=1$, the chain is little different from the previous one, namely, on
average for every $2n$ characters there is a $00$ and a $11$, for
example:

\begin{equation}
\label{0110100}010101011001010101010...
\end{equation}

At $k=n-1$ we get a chain in which for $2n$ symbols there is on
average one pair $01$ and one pair $10$, the rest are $00$ and $11$.
At $n\gg1$ such chain approaches the deterministic chain in which
$n$ zeros is followed by $n$ ones:

\begin{equation}
\label{0..01..1}00000...011111...1
\end{equation}

Let us denominate chain (\ref{0..01..1}) as $S_-$ and call it the
zeros and ones separation chain. This pertains to both open and
closed chains. In the latter case, it is of no importance whether
all zeros are located on the left or on the right.

Let us point out here, that a sequence like (\ref{0..01..1}) can be
obtained by means of the Markov's chain of order $n$. Examining the
deterministic chain, we label the transition probabilities with $0$
or $1$ values only. Such probabilities result in generating of
periodic sequences with different periods. The smallest period
contains one character, the largest -  $2n$ symbols. It  is
therefore possible to distribute the  transition probabilities in
such a way as to generate  a strictly  periodic chain, in which $n$
zeros should be followed by $n$ ones.  Generally speaking,  any of
the periodic sequences may  be regarded  as an example of ideal
order. If we restrict  ourselves to the Markov's chain of the  first
order and the case with a uniform distribution of zeros and ones,
then let us consider three almost ideal orders, corresponding to
minimum information  and described by formulas (\ref{Hmin}) -
(\ref{H2}).

 Order (\ref{0110100}) is only a minor violation
of order (\ref{0101}). However, in transition from $k = 0$ to $k
=1$, i.e. with minimum periodicity disruption, the quantity of
information leaps infinitively within $n\rightarrow\infty$.
 From the information theory viewpoint,
this is not surprising - it is the same as the infinite information
of a black point on a white background.  But when  describing  a
real physical object such result  would appear  unsatisfactory.

 The sequences of ideal order  (\ref{0101}) and (\ref{0..01..1}),
obtained within the framework of the discussed problem, are examined
below as order poles.

\section{Determination of  order poles and parameter  based upon
 the connection function }

\label{part3}
 Let us define the function of connection between two
adjacent symbols $u_{i,i+1}$. Such function may have a real sense,
for example, the  binding  energy between pairs of complementary
nucleotides  in the DNA. Not so long ago, these energies have been
measured   ~\cite {art9}. Let us build a sum, properties  of which
relate  to the idea of order poles and parameter.

 \begin{equation}
\label{Sepsil} F=\sum_{i=1}^{2n}u_{i,i+1},
\end{equation}
where in case of a closed circular chain we have
$u_{2n,2n+1}=u_{2n,1}$, whereas for an open chain, adding up in
(\ref{Sepsil}) exists only until $2n-1$.

For  a chain,  consisting of zeros and ones, the  connection
function  argument  for its next-door neighbors are four possible
sequences $11$, $10$ ,$01$, $00$. Accordingly, we'll define 4 values
for $u_{i,i+1}$: $\varepsilon_{11}$, $\varepsilon_{10}$
,$\varepsilon_{01}$, $\varepsilon_{00}$.  For the sake of  symmetry
let us put  $\varepsilon_{10}=\varepsilon_{01}$ . Let us break  a
chain  into blocks. The block here and  below is a sequence of
identical  symbols, each of them surrounded  by another symbol from
both sides. Let the number of blocks of one of the symbols be equal
to $b$. It is clear that  the number of blocks of the second
character and the number of $01$ or $10$ pairs in the chain also
equal $b$. Since there  is an equal number of zeros and  ones in the
chain, the number  of $00$  and  $11$ pairs in it equals $k=n-b$.
Then,  from (\ref{Sepsil}) it should be:

\begin{equation}
\label{Sb}
F=n\left(\varepsilon_{11}+\varepsilon_{00}\right)+b\Delta\epsilon
\end{equation}

\begin{equation}
\label{deltaeps}\Delta\epsilon=2\varepsilon_{10}-\left(\varepsilon_{11}+
\varepsilon_{00}\right)
\end{equation}

For the long chains, the entered  total function  of connections  of
the next-door  neighbors (\ref{Sepsil}) by structure equals result
(\ref{H2n}), obtained based upon the first-order Markov chain's
model. Conformity of formulas (\ref{Sb}) and (\ref{H2n}) is reached
assuming that

\begin{equation}
\label{eps1100}
\varepsilon_{11}=\varepsilon_{00}=-\log_2\frac{n-b}{n}
\end{equation}

\begin{equation}
\label{eps10} \varepsilon_{10}=-\log_2\frac{b}{n}
\end{equation}

Assuming that the order formula, based on a probabilistic  model,
corresponds to the  general structure (\ref{Sepsil}), we can try to
determine the  order parameter, based on the  latter.  Unlike
probabilistic model, where $u_{i,i+1}$  are determined statistically
and depend on the state  of the entire chain,  we assume  that
parameters  $\varepsilon_{11}$, $\varepsilon_{10}$
,$\varepsilon_{01}$  are constant, which is justified, because the
adjacent elements  of the  chain are "unaware"  of  the chain's
general status. First and foremost, let us find the extremums of
function (\ref{Sepsil}), being limited to the closed chain. In two
cases, when $F(b)$ differs from the constant,  we have:

\begin{equation}
\Delta\epsilon<0 \Rightarrow F_{max}=F(0),\, F_{min}=F(n)
\end{equation}
\begin{equation}
\Delta\epsilon>0\Rightarrow F_{max}=F(n),\, F_{min}=F(0)
\end{equation}

Thus, the state of ideal periodicity $b=n$ and the state of complete
separation $b=0$  (disintegration on two) appear to be very specific
indeed - in these states either minimum or maximum of function
(\ref{Sepsil}) is achieved.

Function  (\ref{Sepsil}) has an additivity property almost relative
to the cross-linking operation of two chains $A$ and $B$ (not
necessarily with the equal number of zeros and ones):

\begin{equation}
\label{kvaziadd} F(A\_B)=F(A)+F(B)+\gamma
\end{equation}
where $A\_B$  denotes the  chain cross-linked together,  $\gamma$ is
a parameter that depends on symbols in the cross-link point .  If
two chains  with open ends are cross-linked, then $\gamma=u_{i,i+1}$
, where $i,i+1$ are numbers  of the cross-linked elements in chain
$A\_B$. It is also possible to cross-link chains, closed in a ring.
For this purpose let us cut each in some place and cross-link the
ends. If, when cross-linked,  all four elements  are identical,  or
if one of $A$ bonds  with zero of $B$, and one of $B$ bonds  with
zero of $A$, then $\gamma=0$. If two zeros of $A$ bond with two ones
of $B$, then  $\gamma=\Delta\epsilon$ . If zero of $A$ bonds with
zero of $B$, and one of $A$ bonds with one of $B$, then
$\gamma=-\Delta\epsilon$ . In order to  introduce the order
parameter, you must  determine  what properties  it should possess.
We can act by analogy  with  the way in the classic theory a formula
for the Shannon's information is being derived from the axioms. May
we remind that  one of the axioms is the information additivity. The
role of similar axiom in our case will be played by the
"near-additivity" property (\ref{kvaziadd}). A set of functions,
defined on closed chains $A_n$  with equal numbers of zeros and ones
and possessing property  (\ref{kvaziadd}), can be represented  as:

\begin{equation}
\label{fAn} f(A_n)=Cn+b\Delta\epsilon
\end{equation}
where $C$  is an  arbitrary constant.  Let us define the order
parameter $K(b,n)$ in the form of renormalized function (\ref{fAn}):

\begin{equation}
\label{Kbn} K(b,n)=\beta f(A_n)
\end{equation}

We want the order parameter to possess certain  symmetry  at the
poles. Let us note that the symmetry requirement $K(0,n)=K(n,n)$ is
insoluble for the $C$ constant, it is therefore natural to demand
the antisymmetry property of on the poles:

\begin{equation}
\label{Kbnantisim} K(0,n)=-K(n,n)
\end{equation}

Determining  the   $C$ constant    in accordance  with
(\ref{Kbnantisim})  and choosing $\beta$  for considerations of
convenient normalization,  we obtain:

\begin{equation}
\label{Kbnnorm} K(b,n)=2b-n
\end{equation}

For an ideally periodic chain we get $K=n$, in case of separation of
zeros and ones  $K=2-n$ and at disintegration into two $K=-n$. As we
shall see below, result (\ref{Kbnnorm}) is a particular case of
another order parameter determination.

\section{ Basic idea. Measure of  order for  a  finite sequence,
 containing equal  number of zeros and ones.}

Classic definitions of information amount  are based on a postulate:
there is a mechanism of symbol sequence generation. In the Shannon's
model there  must  be an apparatus that  generates  characters  with
certain  probabilities. The Kolmogorov's complexity ~\cite {art10}
is determined by existence of the algorithm, which can create a
specified sequence, and the amount of information depends on the
algorithm proper. It is clear that if some sequence of symbols is
set, then without defining its generation mechanism or research by
some procedures  it is impossible to define what we would like to
call the amount of information or degree of a symbol sequence order.
We shall proceed from the idea whereby instead of a generating
mechanism  a transformation mechanism is being used, which does not
change the chain composition. The  idea of comparing binary words by
means of transformation is known from works  ~\cite {Damerau},
~\cite {Levenstain}. Let us consider a set  of sequences,  in which
there  are  $n$  ones and  $n$ zeros. All such sequences differ in
the location of symbols, we may call the $S_i$ state, where $i$ is a
state index. Under transformation we understand  transition of
sequence from $S_i$ to $S_j$ by means of transfer  of one or several
adjacent symbols from one part  of sequence to the other.  Let us
call each such transition a step. We believe that transitions should
be executed under certain rules, which we hereinafter call the
ordering method $\Omega$. Let us define following requirements to
$\Omega$:

\begin{itemize}
\item The ordering method must determine what kind of symbol transfers are allowed.

\item
The ordering method must allow transition from any state $S_i$ to
any state $S_j$ in a certain number of steps. The minimum number of
steps required for this  transition, we hereby denote as
$T(S_i,S_j,\Omega)=t_{ij}$. For symmetry sake ,  we also require
$t_{ij}=t_{ji}$ .

\end{itemize}

The examined system assumes introduction of order poles $S_+$
(\ref{0101}) and $S_-$ (\ref{0..01..1}) in accordance with the above
concept. Indeed, $S_+$ is maximum and $S_-$ is minimum in terms of
number of $01$ and $10$ pairs.  Relative complexity of the $S_+$ and
$S_-$ poles appears maximum  for the $\Omega$ rules, considered
below, i.e. the minimum number of steps necessary for transition
between poles, is no  less than minimum number  of steps for
transition between any other two states.

\begin{equation}
\label{tpolus} t=T(S_+,S_-,\Omega) \geq T(S_i,S_j,\Omega)
\end{equation}
(in (\ref{tpolus}) $S_i$ or  $S_j$ not pole ).

Now we want to build a function, which will compare every state with
a number, the meaning of which is degree of the symbol chain
ordering. Since we have identified two states of the ideal order
$S_+$ and $S_-$ , an arbitrary state of  $S_i$ according to the
general idea is characterized  by two parameters, determining  the
number of steps you need to reach the poles from the $S_i$ state
applying the $\Omega$ rules.

\begin{equation}
\label{tminplus} t_{i+}=T(S_i,S_+,\Omega) \qquad
t_{i-}=T(S_i,S_-,\Omega)
\end{equation}

Let us look for the  order  parameter $K(S_i)$ as a function
$K(t_{i+},t_{i-})$ that satisfies the  next  properties (for brevity
sake, dependence on $\Omega$ in certain places is being ignored):

\begin{itemize}
\item

First, unlike properties of classic entropy, we would like small
changes in a sequence to always result in small changes of the order
parameter. We shall formulate a stricter requirement. Let one step
to have been accomplished according to the $\Omega$ rule and the
sequence transferred from state $S_i$  to $S_j$ . In the new state
we have an order parameter $K(t_{j+},t_{j-})$. If the sequence is
long enough, then a one-step transition is local, i.e. changing
order in small part of the sequence only. It is therefore  natural
to require that  change in the order parameter  under such
transition from any state  would not  depend on the arrangement  of
all sequence in general, i.e. would not depend on $t,\,t_{i+}$ and
$t_{i-}$ , but  exclusively on $\Delta t_+=t_{j+}-t_{i+}$  and
$\Delta t_-=t_{j-}-t_{i-}$:

\begin{equation}
\label{deltaK}
 K(t_{j+},t_{j-})-K(t_{i+},t_{i-})=C(\Delta t_+,\Delta t_-)
\end{equation}

\item The second requirement is the anti-symmetry of the poles' values:

\begin{equation}
\label{AntisimmK}
 K(0,t)=-K(t,0)
\end{equation}
\end{itemize}

(Should we require symmetry of the order parameter on the poles
instead of anti-symmetry (\ref{AntisimmK}), then  from
(\ref{deltaK}) we get $K\equiv Const$.)

Property (\ref{deltaK}) is the necessary and sufficient condition
for linearity of function $K(t_{i+},t_{i-})$, whereas consideration
of (\ref{AntisimmK}) allows us to build a formula for the
calculation of the order parameter.

\emph{Necessity.} Suppose  function $K(t_{i+},t_{i-})$ is linear.
\begin{equation}
\label{LinearK}
 K(t_{i+},t_{i-})= \alpha t_{i+}+\beta t_{i-}+c,
\end{equation}
where values   $\alpha$, $\beta$, $c$ - are constants. Then :
\begin{equation}
\label{deltaKt}
 K(t_{j+},t_{j-})-K(t_{i+},t_{i-})=\alpha \Delta t_++\beta\Delta
 t_-,
\end{equation}
at that,  due to minimality of  $t_{i+},t_{i-},t_{j+},t_{j-}$ ,
variables $\Delta t_+,\Delta t_-$  can only have values of 0,1,-1.

\emph{Sufficiency.} Suppose difference
$K(t_{j+},t_{j-})-K(t_{i+},t_{i-})$ is independent of
$t_{i+},\,t_{i-}$  for all $i$ states. In particular, it should be
true for all states the chain passes along its shortest way from
$S_+$ to $S_-$ . For every such transition we have:

\begin{equation}
\label{tjplusminus}
 t_{j+}=t_{i+}+1\qquad t_{j-}=t_{i-}-1
\end{equation}
 Then, during this transition

\begin{equation}
\label{KminusKeqa}
 K(t_{j+},t_{j-})-K(t_{i+},t_{i-})=a,
\end{equation}
where $a$ is constant.  From (\ref{tjplusminus}) and
(\ref{KminusKeqa}) we get:
\begin{equation}
\label{aeq}
 a=\frac{1}{t}(K(t,0)-K(0,t))
\end{equation}
\begin{equation}
\label{Ktjeq}
 K(t_{i+},t_{i-})=\frac{1}{t}(K(0,t)t_{i-}+K(t,0)t_{i+})
\end{equation}

From the locality requirement it follows that  $a$ in (\ref{aeq})
does not depend on $t$. Using the  antisymmetry requirement  for the
poles' order parameter and choosing a suitable scale, we get a
formula to calculate the order parameter:

\begin{equation}
\label{Ktjeqfin}
 K(t_{i+},t_{i-})=t_{i-}-t_{i+}
\end{equation}

Now enter the relative order parameter of the $S$ chain.

 \begin{equation}
\label {krelative} k(S,\Omega)=\frac{ K(S,\Omega)}{t(\Omega)}
\end{equation}
Then for any ordering method, we get:
\begin{equation}
\label {kS+S-} k(S_+,\Omega)=1,\quad k(S_-,\Omega)=-1
\end{equation}
Enter determination of chaos. We talk about the state of chaos, if:
\begin{equation}
\label {k=0} k(S,\Omega)=0
\end{equation}
or about  the near-chaos state,  if:
\begin{equation}
\label {k=0} \vert k(S,\Omega)\vert\ll1
\end{equation}

Let us also define a value we call $\Omega$-information in the $S$
chain:

\begin{equation}
\label {Inf} I=I(S,\Omega)=\left\{\begin {array}{ll}
T(S,S_+,\Omega), & K(S,\Omega)\geq0\\T(S,S_-,\Omega), &
K(S,\Omega)<0\end{array}\right.
\end{equation}

That is, the amount of $\Omega$-information equals minimum number of
transfers for transition to the nearest pole at a given $\Omega$.
Note, that the absolute order parameter and the $\Omega$-information
are not additive values in terms of bonding operations:

\begin{equation}
\label{notaddit} K(A\_B,\Omega) \neq K(A,\Omega)+K(B,\Omega)
\end{equation}

\section{The order parameter for some methods of transfer.}

Considering the various ways of transfer further, we shall be
interested  in finding such symbol location conditions that lead to
the state  of chaos. Another  issue  is to find such $\Omega$, for
which condition of the order parameter's additivity is met with
respect to chain bonding operations. It is also interesting to see
how the order parameter and the amount  of $\Omega$-information will
change with a small chain deviation from its ideal order.

 \subsection{Method  of a single symbol arbitrary transfer.}
 \label{sec:OneTrasport}

Let us assume it is allowed to take any symbol and to transfer it by
inserting between any two symbols or to  the end of the line. Let us
denote  such sequencing  method  as $\Omega_1$. The  number  of
transfers required  for transition between the poles (\ref{0101}),
(\ref{0..01..1}) by means of $\Omega_1$ is:

\begin{equation}
\label{t(Omeg1} t(\Omega_1)=n-1
\end{equation}

It is easy to verify that (\ref{tpolus}) holds, and in the case of
closed chains, it is strict. For the poles (\ref{0101}),
(\ref{0..01..1}) we get:

\begin{equation}
\label{KS+KS-} K(S_+)=n-1,\quad K(S_-)=-n+1
\end{equation}
For brevity, in this section we present results for open chains
only.

\paragraph {\indent Small deviation from the ideal order.}

Let us consider chain $S_1$ , near to periodic, similar to
(\ref{0110100}), in which there  are $2n-2$ pairs of $01$ or $10$,
and once we have both $00$ and $11$. (Although we discuss an open
chain, when counting pairs it is more convenient to regard the chain
as closed in the ring.)

\begin{equation}
\label{S1} T(S_1,S_+)=1=I,\quad T(S_1,S_-)=n-i, \quad K(S_1)=n-i-1
\end{equation}

In (\ref{S1}) $i=1$, in case $00$ and $11$ are inside the chain and
$i=2$, if $00$ or $11$ are on its edge. The relative order parameter
for the examined chain:
 \begin{equation}
\label{kS1} k(S_1)=1-\frac{i}{n-1}
\end{equation}
Note also that
\begin{equation}
\label{KkS+S1} K(S_+)-K(S_1)=i,\quad k(S_+)-k(S_1)=\frac{i}{n-1}
\end{equation}

Thus,  unlike the  Shannon's  entropy,  a  small  deviation of an
arbitrarily long chain from the perfectly  periodic one causes a
finite change of the  order parameter and the $\Omega$-information
amount, whereas difference in relative order parameters tends  to
zero.

Let us consider a near-separation chain (\ref{0..01..1}).  We assume
one of the ones has been transferred  and  introduced  between
zeros. Let us denote such chain as $S_2$. Then:

\begin{equation}
\label{kS2} K(S_2)=3-n, \quad K(S_-)-K(S_2)=-2.
\end{equation}
In this case, for an arbitrarily long chain the change of the
absolute order parameter is finite as well, whereas the amount of
$\Omega$-information in such chain is  $I(S_2)=1$.

 \paragraph {\indent  The order parameter calculation formula.}

 Let us give an  example of calculation of the order parameter and the $\Omega$-information
  amount for the chain $S$ in the figure below. Upper pointers show transfer to the state
  of separation, lower ones - to the state of ideal periodicity.

\begin{figure}
\begin{center}
\includegraphics[angle=90, height=0.5\textwidth]{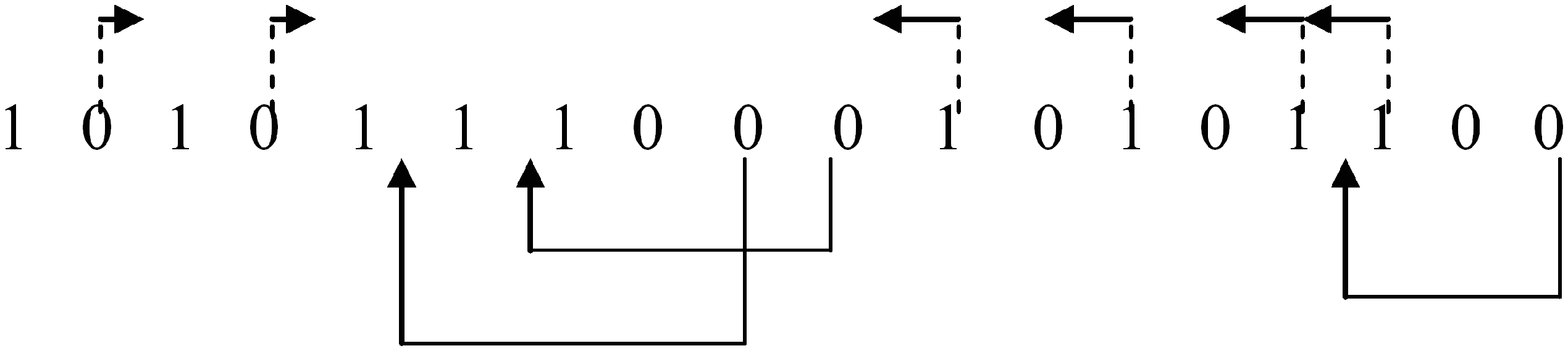}
\end{center}
\end{figure}

\begin{displaymath}
 T(S,S_+)=3=I(S), \quad T(S,S_-)=6, \quad K(S)=3, \quad k(S)=\frac{3}{8}
\end{displaymath}

It is clear that $T(S,S_+)$ always equals the  number  of intervals
between consecutive ones or zeros (for exact  calculation  the chain
should be closed  into a ring). Determination of $T(S,S_-)$ is not
that obvious. Let us break up the chain into blocks, number the
blocks from left to right by means of the $i$ index and denote  the
$s_i$ number  of symbols in a block as ($s_i\geq 1$). Let us count
the number of zeros and ones to the block's left. Let $l_{si}$  be
the number of symbols to the block's left (same as in the block
proper), and $\bar{l}_{si}$  - the  number  of other  symbols to the
left. Then, to the block's right there are $n-s_i-l_{si}$  symbols
of the block and $n-\bar{l}_{si}$  of other symbols. In order to
pass to the state of complete separation it is necessary to transfer
symbols of one kind (on the block's left) to the right, and symbols
of other kind (on the block's right) to the left. We get two
possible numbers of transitions:

\begin{equation}
\label{x1} x_1=\bar{l}_{si}+n-s_i-l_{si}
\end{equation}
\begin{equation}
\label{x2} x_2=n+l_{si}-\bar{l}_{si}
\end{equation}
Let us choose the smallest of the two variables $x_1,\,x_2$   by
means of the function:

\begin{displaymath}
\label{minx1_x2} f_i=\min (x_1,x_2)=\left\{\begin{array}{ll} n+l_{si}-\bar{l}_{si} & s_i\leq2(\bar{l}_{si}-l_{si})\\
\bar{l}_{si}+n-s_i-l_{si} & s_i>2(\bar{l}_{si}-l_{si})
\end{array}\right.
\end{displaymath}

 Then: $$T(S,S_-)=\min_i(f_i)$$

A few additional examples of calculation of the order parameter and
$\Omega$-information.

$$A: \,\,011111010000010011$$
$$T(A,A_-)=5,\quad T(A,A_+)=5,\quad K(A)=0,\quad I(A)=5,$$
$$B: \,\,11101011100000$$
$$T(B,B_-)=2,\quad T(B,B_+)=4,\quad K(B)=-2,\quad I(B)=2.$$

\paragraph {\indent The "bonding" operation and discussion of additivity.}
Although operation of cross-linking chains $A$ and $B$ is not
additive, as we have noted already (\ref{notaddit}), it is possible
to discuss the near-additivity property.

\begin{equation}
\label{Addit} K(A\_B)=K(A)+K(B)+s,
\end{equation}
where $s$ is a small number compared with each of the chains'
length. Clearly, for the near-periodicity chains (\ref{0101}),
condition (\ref{Addit}) holds. The additivity is quintessentially
violated for the near-separation chains. Let us consider bounding of
two type (\ref{0..01..1}) chains. Let chains $A$  and $B$ have
length of $2m$ and $2k$ correspondingly and $k\geq m$:

$$A:\underbrace{ 111...11000...00}_{2m},\, T(A,A_+)=m-1,\, T(A,A_-)=0,\, K(A)=1-m,$$
$$B: \underbrace{ 111...11000...00}_{2k},\, T(B,B_+)=k-1,\, T(B,B_-)=0,\, K(B)=1-k,$$
$$C=A\_B: \underbrace{ 111...11000...00}_{2m}\underbrace{ 111...11000...00}_{2k},$$
$$ T(C,C_+)=m+k-2,\, T(C,C_-)=m,\, K(C)=-k+2. $$

 In the studied case:
\begin{equation}
\label{Addit1} K(A\_B)=K(A)+K(B)+m,
\end{equation}
\paragraph {\indent The "bonding" operation and struggle of orders.}

Let us consider the bonding operation of $n$ identical $A$ chains,
each in the state of complete separation ($s$ of ones followed by
$s$ of zeros ($s>1$)):

$$A:\underbrace{ 111...11000...00}_{2s},\quad  k(A)=-1.$$
Let us call the resulting chain $nA$:

\begin{equation}
\label{nA}\underbrace{
A\_A\_A...\_A}_{n}=nA=\underbrace{\underbrace{
1...110...00}_{2s}\underbrace{ 1...110...00}_{2s}...\underbrace{
1...110...00}_{2s}}_{n}
\end{equation}
It easily to calculate, that:
$$T(nA,(nA)_+)=n(s-1)$$
$$T(nA,(nA)_-)=s(n-1)$$
The absolute order parameter equals:

\begin{equation}
\label{KnA} K(nA)=n-s,
\end{equation}
$\Omega$-information:
\begin{equation}
\label{InA} I(nA)=\left\{\begin{array}{ll}ns-n,& n \geq s\\
ns-s, & n<s\end{array}\right.
\end{equation}
The relative order parameter equals:

\begin{equation}
\label{kknA} k(nA)=\frac{n-s}{ns-1},
\end{equation}
At that
\begin{equation}
\label{LimknA} \lim_{n \rightarrow  \infty}k(nA)=\frac{1}{s},
\end{equation}

Results (\ref{kknA}) - (\ref{LimknA}) describe the struggle of
orders. At $n=1$, we have the order of separation. With $n$
increasing, the relative order parameter increases, passes a
near-chaos point, changes its sign when the chain periodicity begins
to dominate and in the extreme reaches a value, determined  solely
by the period composition.

 A more general result for the struggle of
orders can be obtained if we consider the bounding operation of $n$
identical chains $B$, $2s$ long and with equal number of zeros and
ones, for each of which:

\begin{equation}
\label{TBB+} T(B,B_+)=m \leq s-1
\end{equation}

We obtain:
\begin{equation}
\label{TnBnB+} T(nB,(nB_+))=nm
\end{equation}
\begin{equation}
\label{TnBnB-} T(nB,(nB_-))=sn-l
\end{equation}
(where $l\leq s$).
\begin{equation}
\label{KnB} K(nB)=n(s-m)-l
\end{equation}
\begin{equation}
\label{limknB} \lim_{n \rightarrow  \infty}k(nB)=1-\frac{m}{s}
\end{equation}

 \subsection{ Method  of transfer by blocks.}
\label{subsec:blok}

Let us set the ordering rule $\Omega_2$ that allows transferring the
entire block or its part as a whole. It is easily understood that
$t(\Omega_2)=t(\Omega_1)$. The  calculation rule for $T(S,S_+)$ is
the same, as in $\Omega_1$, however it is different for $T(S,S_-)$.
For  an open chain when it begins and  ends with the same symbol:

\begin{equation}
\label{TSS-b0} T(S,S_-)=b,
\end{equation}
If the ends have different symbols:

\begin{equation}
\label{TSS-b1} T(S,S_-)=b-1.
\end{equation}

If the chain is closed, we have (\ref{TSS-b1}) and in all cases
$T(S,S_+)=n-b$. The  order parameter  for the closed  chain :

\begin{equation}
\label{Ktrasportb} K(S)=2b-n-1.
\end{equation}

Up to an irrelevant unit, formula (\ref{Ktrasportb}) corresponds to
result (\ref{Kbnnorm}), obtained earlier in Section (\ref{part3}).
Therefore, for $\Omega_2$  the order parameter has the property of
additivity almost under the bonding operation (\ref{Addit}), where
$s\in\{0,\pm 1\}$. $\Omega$-information has this property only when
$K(A)$ and $K(B)$ have the  same sign or equal zero.

\begin{equation}
\label{KvaziAdd} K(A)K(B)\geq 0\Rightarrow
I(A\_B)=I(A)+I(B)+s,\,\,s\in\{0,\pm 1\}
\end{equation}

If $K(A)$ and $K(B)$  have different signs, the $A\_B$ cross-link
generates  new $\Omega$-information. For example, if we take two
chains of equal length  $2n$, one of which is in the state  of
separation and the other one is perfectly periodic, the
$\Omega$-information in each of them equals zero, but after bonding
we obtain  $I=n$.

Both for the absolute order parameter and the amount of
$\Omega$-information, the property of near-commutativity holds:

\begin{equation}
\label{KommutK} K(A\_B)=K(B\_A)+s
\end{equation}
\begin{equation}
\label{KommutI} I(A\_B)=I(B\_A)+s
\end{equation}
where, as above,  $s\in\{0,\pm 1\}$.

 Ordering with blocks has one more important property,
  which we formulate as the chaos theorem.

\paragraph {\indent Theorem of  $\Omega_2$ chaos.}

From (\ref{TSS-b0})-(\ref{Ktrasportb}) it follows that for the
$\Omega_2$ rules the state  of chaos occurs when $n=2b$ or $n=2b-1$.

\emph{ Hence,  for the  chain  in question the state of chaos in
relation to transfer  by blocks  is  achieved under the same
conditions that exist when the Shannon's information maximum is
reached for the first-order Markov's chain,  i.e. when frequencies
of $00$ and $01$ pairs' occurrence are equal.}

\paragraph {\indent Struggle of orders. }
Let us calculate  the $\Omega_2$ order parameter  for chain
(\ref{nA}).

We calculate the order parameter $\Omega_2$ for the chain
(\ref{nA}).

$$T(nA,(nA)_+)=n(s-1)$$
$$T(nA,(nA)_-)=n-1$$

\begin{equation}
\label{KnAO2} K(nA)=2n-1-sn,
\end{equation}

\begin{equation}
\label{knAO2} k(s,n)=k(nA)=\frac{2n-1-sn}{ns-1},
\end{equation}

\begin{equation}
\label{LimknAO2} \lim_{n \rightarrow  \infty}k(nA)=\frac{2}{s}-1,
\end{equation}
Similar to the case of a single symbol transfer, in the $\Omega_2$
model the struggle of orders is also observed during cross-linking
of identical sites with separation (\ref{knAO2})-(\ref{LimknAO2}).
At  $s=1$, regardless of $n$,  we obtain  the  ideal periodicity
state,  for $s\rightarrow \infty$ we obtain the state of separation.
The near-chaos state  is reached already at $s=2$. Difference  from
a single symbol transfer is clear - transfer by blocks shortens the
path to the state of complete separation.

\subsection{The adjacent elements' permutation  method  for the $\Omega_3$ ring chains.}

We shall consider closed chains with the equal number of ones and
zeros. (We shall write these chains in a row and assume that the
last element is joined to the first). Note, that in this $S_-$
(\ref{0..01..1}) is equivalent, for example, to the chain like:

$$\underbrace{ 00...0}_{r}\underbrace{ 11...1}_{n}
\underbrace{ 00...0}_{n-r}$$

We shall introduce the $\Omega_3$ ordering rule, which allows
swapping two adjacent characters. In this case, for the chains in
question $(n\geq 2)$:

\begin{equation}
\label{tom3} t(\Omega_3)=T(S_+,S_-)=T(S_-,S_+)=\left\{\begin{array}{ll}0.25n^2,& n =2k,\\
0.25(n^2-1), & n =2k-1,\end{array}\right.
\end{equation}
where $k$ is a natural number. (To get the minimum number  of steps
required for transition from (\ref{0101}) to (\ref{0..01..1}), zeros
should be transferred simultaneously from two opposite sides of the
ones' block).

First of all, let us investigate sequence (\ref{0110100})\,  denoted
below as $G$. To do this we need an additional  parameter $l$,
denoting the minimal  number  of characters  in a circle between
$00$ and $11$ pairs. It is obvious, that $l$ is even, at that $l\leq
n-2$. The number of steps we require to return  the  sequence to a
perfectly periodic state  is:

\begin{equation}
\label{TGG+} T(G,G_+)=0.5l+1
\end{equation}

The number of transfers  for the transition to a complete separation
state:

\begin{equation}
\label{TGG-} T(G,G_-)=t(\Omega_3)-0.5l-1
\end{equation}

 From (\ref{TGG+}) -(\ref{TGG-}) and definition (\ref{Ktjeqfin})
 we have:

\begin{equation}
\label{KG} K(G)=t(\Omega_3)-l-2
\end{equation}

\begin{equation}
\label{kG} k(G)=1-\frac{l+2}{t(\Omega_3)}
\end{equation}

\begin{equation}
\label{KG+-Kg} K(G_+)-K(G)=l+2
\end{equation}

\begin{equation}
\label{Gl} I(G,l)=0.5l+1
\end{equation}

\begin{equation}
\label{kG+-kg} k(G_+)-k(G)=\frac{l+2}{t(\Omega_3)}
\end{equation}

Change of the relative order parameter, as before, tends to zero at
$n \rightarrow \infty$ and  constant $l$. However, change of the
absolute  order parameter and  the amount of $\Omega$-information
increases linearly with increasing $l$. In particular, if $00$ and
$11$ divide the chain into two equal parts, then $l=n-2$ and hence:

\begin{equation}
\label{Gn} I(G,n-2)=0.5n
\end{equation}

Let us analyze  the  previously examined  chain,  obtained  from the
state of separation through  transfer  of one of the  ones in a
zeros' block.

 \begin{equation}
\label {00010111} E: \underbrace{ 000010...0}_{n+1}\underbrace{
1111...1}_{n-1}
\end{equation}

Suppose this one is located l zeros from the nearest interface of
zeros and ones. Then:

 \begin{equation}
\label{TEE-} T(E,E_-)=l,\quad T(E,E_+)=t(\Omega_3)-l
\end{equation}

\begin{equation}
\label{kE} K(E)=2l-{t(\Omega_3)},\quad k(E)=\frac{2l}{t(\Omega_3)}-1
\end{equation}

\begin{equation}
\label{DK} K(E)-K(E_-)=2l
\end{equation}

\begin{equation}
\label{Dk} k(E_)-k(E_-)=\frac{2l}{t(\Omega_3)}
\end{equation}

\begin{equation}
\label{IE} I(E,l)=l
\end{equation}

General conclusions, made for chain $G$, are also true for $E$.

By specific examples one can easily verify that the near-additivity
property is not met in relation to $\Omega_3$ . To verify struggle
of orders let us calculate the  order parameter for the chain
(\ref{nA}) assuming it is ring-shaped. Below we use the notion of
$t(\Omega,j)$  - number of transitions from the ideal periodicity
under the state of separation for a $2j$-long chain. For the
adjacent symbols' permutation method we get:

\begin{equation}
\label{knA}
k(nA)=\frac{s^2t(\Omega_3,n)-nt(\Omega_3,s)}{t(\Omega_3,ns)}
\end{equation}
или:
\begin{equation}
\label{knA1} k(nA)=\left\{\begin {array}{lll}
1-\displaystyle\frac{1}{n}, & s=2i,
& n=2m,\\
1-\displaystyle\frac{s^2-1}{s^2n}, & s=2i-1, & n=2m,\\
1-\displaystyle\frac{1}{n}-\displaystyle\frac{1}{n^2}, & s=2i, & n=2m-1,\\

1+\displaystyle\frac{(1-s^2)(1+n)}{s^2n^2-1}, & s=2i-1, & n=2m-1
\end{array}\right.
\end{equation}

Here, if s = 1, then $n \geq 2$, for the rest of $s$ we have
 $n\geq1$.

Result (\ref{knA1}), generally speaking, is weakly dependent on $s$.
In case \\ $n=1, s \geq 2$, we have the state of separation: $k=-1$.
For $n=2$ and any $s$, we already  have $k>0$, i.e. we skip the
state of chaos, and  with  increasing $n$ we have $k\rightarrow 1$.
Thus, for large $n$ in terms of the  order parameter the  chain
behaves as an ideally periodic one (\ref{0101}).

\section{Conclusion}
Development of the suggested idea is subject to the following
problems' solving:

\begin{itemize}
 \item Generalization of the order parameter determination
 in case of unequal number of zeros and ones.
\item  Expansion of theory in case of any number of symbols in an alphabet.
 \item Establishing criteria  for distinguishing  one ordering method
  from another and selection of optimal  ordering method.
\end{itemize}
\section{Acknowledgment}

 The author expresses his deep gratitude to Professor Eduard Yakubov and Dr.
 Dmitry Goldstein for the opportunity to present the work
 at the mathematical colloquium at the Holon Institute of Technology and discussion of results.

\end{document}